\documentstyle[12pt]{article}

\newcommand{\be}{\begin{equation}}
\newcommand{\ee}{\end{equation}}
\newcommand{\bea}{\begin{eqnarray}}
\newcommand{\eea}{\end{eqnarray}}
\newcommand{\gsim}{ \mathop{}_{\textstyle \sim}^{\textstyle >} }

\def\lm{\lambda}

\def\Sig{\Sigma}

\def\pdy{\partial_y}
\def\vev#1{\langle #1 \rangle}

\newcommand\bh{\bar{H}}
\newcommand\hc{H^c}
\newcommand\bhc{\bar{H}^c}

\newcommand\bphi{\bar{\phi}}
\newcommand\phic{\phi^c}
\newcommand\bphic{\bar{\phi}^c}

\def\bx{\bar{X}}
\def\bm{\bar{M}}
\def\tr{{\rm tr}}

\begin{document}
\setlength{\baselineskip}{0.7cm}

\begin{titlepage}
\null
\begin{flushright}
{\tt hep-ph/0201216}\\
UT-02-04\\
January, 2002
\end{flushright}
\vskip 1cm
\begin{center}
{\Large\bf 
Light Higgs Triplets in Extra Dimensions 
%Is the triplet Higgs really heavy?
}

\lineskip .75em
\vskip 1.5cm

\normalsize

{\large Naoyuki Haba$^{a}$}
%\footnote{e-mail address: haba@eken.phys.nagoya-u.ac.jp} 
and {\large Nobuhito Maru$^{b}$}
%\footnote{e-mail address: maru@hep-th.phys.s.u-tokyo.ac.jp; 
%JSPS Research Fellow.} 

\vspace{1cm}

{\it $^{a}$Faculty of Engneering, Mie University, Tsu, 
Mie, 514-8507, JAPAN} \\
%{\it $^{b}$Department of Physics, Nagoya University, 
%Nagoya, 464-8602, JAPAN} \\
{\it $^{b}$Department of Physics, University of Tokyo, 
Tokyo 113-0033, JAPAN} \\

%\vspace{5mm}

%{\it $^{b}$Research Center for the Early Universe, 
%University of Tokyo \\
%Tokyo 113-0033, JAPAN}
%
\vspace{18mm}

{\bf Abstract}\\[5mm]
{\parbox{13cm}{\hspace{5mm}
%
%%%%%%%%%%%%%%%%%%%%%%%%%%%%%%%%%%%%%%%%%%%%%%%%%%%%%%%%%%%%%%%%%%
We discuss the possibility that 
the Higgs triplet can be light 
(1TeV in the most interesting case) 
without contradictions with the proton stability 
%and the gauge coupling unification 
in the context of higher dimensional theory. 
The proton stability is ensured 
by the suppression of Yukawa coupling of 
the Higgs triplet to the matter 
through its small overlap of wave functions 
in extra dimensions. 
The light Higgs triplets might be detected 
in future collider experiments 
as an alternative signature of GUT 
instead of the proton decay. 
The gauge coupling unification can be preserved 
by introducing extra bulk matter fields. 
%%%%%%%%%%%%%%%%%%%%%%%%%%%%%%%%%%%%%%%%%%%%%%%%%%%%%%%%%%%%%%%%%%%
}}

\end{center}

\end{titlepage}

%%%%%%%%%%%%%%%%%%%%%%%%%%%%%%%%%%%%%%%%%%%%%%%%%%%%%%%%%%%%
%                          Main part                       %
%%%%%%%%%%%%%%%%%%%%%%%%%%%%%%%%%%%%%%%%%%%%%%%%%%%%%%%%%%%%

%\section{Introduction}
One of the serious problems in Grand Unified Theories (GUTs) 
\cite{GUT} is the doublet-triplet splitting problem. 
We have to explain in a natural way 
why the Higgs doublet mass is the weak scale 
and the Higgs triplet mass is at least the GUT scale 
after the GUT symmetry breaking.
\footnote{There has been many proposals for this problem 
\cite{Missing}-\cite{witten}.} 
In usual, 
this mass splitting is explained 
by an unnatural fine-tuning of parameters in the theory. 
%In this argument, 
%Yukawa coupling constants of the Higgs triplet and the matter 
%are assumed to be of order unity 
%as long as there is no reason to be suppressed 
%by the symmetry. 
The Higgs triplets must be superheavy 
(at least the GUT scale $\simeq 10^{16}$ GeV.) 
%since these unsuppressed coulpings cause the rapid proton decay 
otherwise the rapid proton decay is caused 
by the dimension five operators \cite{SYW}.
\footnote{Throughout this paper, R-parity is assumed.} 
If Yukawa coupling constants of 
the Higgs triplet to the matter 
are extremely small, 
the proton decay can be suppressed and 
the Higgs triplets need not to be superheavy. 
The question is whether such a situation 
is naturally possible or not. 
The answer is yes. 
If we consider the extra dimensions and 
the Higgs triplets and the matter are localized 
at different points in extra dimensions, 
the effective Yukawa couplings 
in four dimensions are highly supperessed 
due to the small overlap of wave functions \cite{AS}.

In this paper, 
we apply this mechanism to the doublet-triplet splitting 
and discuss the possibility that the Higgs triplets can be light 
without contradictions with the proton stability 
%and the gauge coupling unification 
in the extra dimensional framework. 
Dvali \cite{dvali} has also proposed a similar scenario 
in four dimensional theory, 
where Yukawa couplings of the Higgs triplet to the matter 
are suppressed for group theoretical reasons. 
However, the complicated superpotential in the Higgs sector 
is required to obtain the desired vacuum expectation value (VEV) 
of the adjont Higgs field. 
In our scenario, 
these couplings are suppressed 
by the dynamics in extra dimensions (see also \cite{KY}.) 
and our model is very simple.

Let us consider a supersymmetric (SUSY) SU(5) GUT 
in five dimensions, for concreteness. 
The model is based on Ref.~\cite{maru1}. 
The action of the Higgs sector is 
\bea
\label{Higgsaction}
&&S = \int d^4xdy \left[ \int d^4 \theta 
(H^{\dagger} e^{-V} H + H^{c{\dagger}} e^V \hc 
+ \bh^{\dagger} e^V \bh 
+ \bar{H}^{c{\dagger}} e^{-V} \bhc ) \right. \nonumber \\ 
&&+ \left\{ \int d^2 \theta 
\left( {\hc ( \pdy + X(y) + M) H 
+ \bhc ( \pdy + \bx(y) + \bm) \bh } \right) \right. \\ 
&&+ \delta(y) 
\int d^2 \theta \left. \left.
\left( \lm_1 {\rm tr}(X^2 \Sig) 
+ \lm_2 {\rm tr}(\bx^2 \Sig) 
+ \lm_3 {\rm tr}(X \Sig^2) 
+ \lm_4 {\rm tr}(\bx \Sig^2) 
+ \frac{1}{2} m_0 {\rm tr}(\Sig^2) \right) 
+ {\rm h.c.} \right\} \right], \nonumber
\label{action}
\eea
where $H (\bh), \hc (\bhc)$ are left-handed 
(charge conjugated right-handed) 
${\cal N} = 1$ SUSY in four dimensional chiral 
superfield components of the single ${\cal N} = 1$ 
SUSY in five dimensional chiral superfield 
$H({\bf 5}) = \left( H, \bh^c \right)$ and 
$\bh ({\bf \bar{5}}) = \left( \bh, H^c \right).$ 
${\bf 5}, {\bf \bar{5}}$ are the corresponding 
representations of SU(5). 
$X(y), \bx(y)$ are the bulk fields in the ${\bf 24}$ dimensional 
representation under $SU(5)$. 
%\footnote{$X(y)$ and $\bx(y)$ are rescaled by $M_*^{-1/2}$, 
%where $M_*$ is the Planck scale in five dimensional theory, 
%since their mass dimension is 3/2 in five dimensions.} 
$\Sig$ is an usual SU(5) GUT adjoint Higgs field, 
which is assumed to be localized on the brane at $y=0$. 
We assume that $X(y),\bx(y)$ depends on $y$, 
and $M,\bm$ do not. 
$\lm_{1\sim4}$ are dimesionless constants and 
$m_0$ is a mass parameter. 
This formulation of the action Eq.~(\ref{Higgsaction}) is useful 
because it is written by using the ${\cal N}=1$ superfield formalism 
and ${\cal N}=1$ SUSY in four dimensions 
is manifest \cite{AHHSW,NAHTGJW}. 
 From F-flatness conditions $\partial W/\partial X = 0$ 
and $\partial W/\partial \bx = 0$, one obtains 
\bea
&& \hc H -\frac{1}{5} \tr (\hc H) = 0,
~\bhc \bh - \frac{1}{5} \tr (\bhc \bh) = 0, \\
\label{f}
&& 2 \lm_1 X(0) + \lm_3 \Sig = 0,
~2 \lm_2 \bx(0) + \lm_4 \Sig = 0. 
\eea
It is remarkable that Eqs.~(\ref{f}) 
connect $\langle X(0) \rangle$ and $\langle \bx(0) \rangle$ 
in the bulk with $\vev \Sig$ on the brane at $y=0$.\footnote{
One might think that the GUT breaking VEV of the bulk field 
can be directly obtained from the minimization of the potential. 
But it is impossible because ${\cal N}=1$ SUSY in five dimensions 
highly constraints the form of the superpotential.} 
Using Eqs.~(\ref{f}), 
the last term in Eq.~(\ref{Higgsaction}) reproduces 
the Higgs superpotential 
%and F-flatness $\partial W/\partial \Sig$ 
%reproduces its stationary condition 
in the minimal SU(5) GUT. 
Expanding 
%the five dimensional superfields 
$H, \hc, \bh$ and $\bhc$ by the mode functions as 
\bea
H(x,y) = \sum_n \phi_n(y) H_n(x),&~&
\hc(x,y) = \sum_n \phic_n(y) \hc_n(x), \\
\bh(x,y) = \sum_n \bphi_n(y) \bh_n(x),&~&
\bhc(x,y) = \sum_n \bphic_n(y) \bhc_n(x), 
\eea
where $x$ denotes the coordinates of 
the four dimensional space-time, 
the equations of motions 
for the zero mode wave functions of Higgs fields are obtained 
\bea
(\pdy + X(y) + M)~\phi_0(y) &=& 0, \\
(- \pdy + X(y) + M )~\phic_0(y) &=& 0, \\
(\pdy + \bx(y) + \bm)~\bphi_0(y) &=& 0, \\
(-\pdy + \bx(y) + \bm)~\bphic_0(y) &=& 0. 
\eea

Assuming that $X(y) = X(0) + a^2 y$, 
$\bx(y) = \bx(0) + a^2 y$ in a small region 
of the point crossing zero, 
where $a$ is a dimensionful constant, 
we obtain two Gaussian normalizable zero mode wave functions
\footnote{The other zero modes should be vanished 
since they are not normalizable.}
\bea
\label{h}
\phi_0(y) &\sim& {\rm exp}\left\{ -\frac{a^2}{2} 
\left(y - \frac{X(0) + M}{a^2} \right)^2 \right\}, \\
%\phic_0(y) &\sim& {\rm exp} \left\{\frac{a^2}{2} 
%\left(y - \frac{X(0) + M}{a^2} \right)^2 \right\}, \\ 
\label{hb}
\bphi_0(y) &\sim& {\rm exp} \left\{ -\frac{a^2}{2} 
\left(y - \frac{\bx(0) + \bm}{a^2} \right)^2 \right\}. 
%\bphic_0(y) &\sim& {\rm exp} \left\{ \frac{a^2}{2} 
%\left(y - \frac{\bx(0) + \bm}{a^2} \right)^2 \right\}. 
\eea

Before discussing the doublet-triplet splitting problem 
in detail, 
we comment on various scales in our model. 
There are three typical mass scales, 
the Planck scale in five dimensions $M_*$, 
the wall\footnote{On this wall, 
the matter and Higgs fields are localized.} 
thickness scale $L^{-1}$, 
which should be considered as the compactification scale 
and the inverse width of Gaussian zero modes $a^{-1}$. 
As explained in Ref. \cite{AS}, 
for the description to make sense, 
the wall thickness $L$ should be larger than 
the inverse width of Gaussian zero modes $a^{-1}$ 
so that the wall has enough width to trap matter and Higgs modes. 
Furthermore, $a^{-1}$ should be smaller than 
or equal to the five dimensional 
Planck length $M_*^{-1}$, 
\be
L^{-1} < a \le M_*. 
\ee
We take $L^{-1}$ to be $M_{GUT}$ in order to preserve 
the gauge coupling unification. 
The five dimensional Planck scale $M_*$ can be taken 
to be about $10^{17}$ GeV or $10^{18}$ GeV. 
Throughout this paper, 
$M_*$ is taken to be $10^{18}$ GeV 
and $a \simeq M_*$ for simplicity.

Now, we propose two scenarios of 
the doublet-triplet splitting problem. 
The first one which realizes the doublet-triplet splitting 
is based on the shining mechanism \cite{AHHSW}. 
%the bulk Higgs mass term. 
We introduce a singlet superfield and consider the overlap 
between the Higgs fields and the singlet field. 
As explained in Ref.~\cite{maru1}, 
the simplest case without a singlet is not realistic 
because the Higgs doublets are too apart from each other 
to yield the hierarchy between the top and the bottom 
Yukawa couplings naturally.
\footnote{Assuming the top Yukawa coupling constants 
in five and four dimensional theories and 
the bottom Yukawa coupling constant in five dimensional theory 
to be of order unity, 
the effective bottom Yukawa coupling constant 
in four dimensional theory becomes smaller 
than ${\cal O}(10^{-21})$. 
In order to explain the observed bottom mass, 
we have to assume an unnatural huge 
bottom Yukawa coupling constant 
in five dimensional theory.} 
%Therefore, 
%we introduce a singlet superfield and consider the overlap 
%between the Higgs fields and the singlet field. 

The action of the singlet sector is 
%based on the ``shining" mechanism \cite{AHHSW} 
%
\be
S = \int d^4x dy \left[ \int d^4\theta (S^{\dagger}S + S^{c\dagger}S^c) 
+ \left\{ \int d^2\theta S^c (\partial_y + m_s) S 
- \delta(y) \int d^2\theta JS^c + {\rm h.c.} \right\} \right], 
\ee
where $S$ is a bulk SU(5) singlet superfield, 
$S^c$ is its conjugated superfield, 
$J$ is a constant and $m_s$ is a mass parameter. 
F-flatness conditions lead to 
\be
S = \theta(y) J e^{-m_s y},~S^c = 0,  
\ee
where $\theta(y)$ is a step function of $y$. 
The doublet-triplet splitting is achieved by the coupling 
\bea
&& \frac{1}{\sqrt{M_*}} \int d^4x dy \left\{ \int d^2 \theta 
S (x,y) H (x,y) \bh (x,y) + {\rm h.c.} \right\} \\
%&& = \frac{1}{\sqrt{M_*}} \int dy S^c (y) \phi_0(y) \bphi_0(y) 
%\int d^4x d^2 \theta H_0(x) \bh_0(x) + {\rm h.c.} \\
%&& = \frac{1}{\sqrt{M_*}} \int dy \theta(-y) J e^{m_s y} 
%\sqrt{\frac{a^2}{2 \pi}} {\rm exp} 
%\left[ -\frac{a^2}{2} \left( y- \frac{X(0) + M}{a^2} \right)^2 
%- \frac{a^2}{2} \left( y - \frac{\bx(0) + \bm}{a^2} \right)^2 \right] 
%\nonumber \\
%&& \int d^4x d^2 \theta H_0(x) \bh_0(x) \\
&& \simeq M_* {\rm exp} 
\left[ - \frac{1}{2M_*^2} \left\{ (X(0) + M)^2 + (\bx(0) + \bm)^2 \right\} 
+ \frac{(X(0) + M + \bx(0) + \bm - m_s)^2}{4M_*^2} \right] \nonumber \\ 
&& \times \int d^4x d^2 \theta H_0(x) \bh_0(x) + {\rm h.c.}, 
\eea
where $J \simeq M_*^{3/2}$ are assumed. 
As mentioned in Ref.~\cite{maru1}, 
an R-symmetry for instance  has to be imposed to forbid 
the bulk Higgs mass term. 
In order for Higgs doublets $H_2$, $\bar{H}_2$ 
to be the weak scale, 
\be
\label{doublet}
M_2 \simeq M_* {\rm exp} 
\left[ -\frac{1}{2} \{(-3x + m)^2 + (-3\bar{x} + \bar{m})^2 \} 
+ \frac{1}{4} (-s - 3x + m - 3\bar{x} + \bar{m})^2 \right] 
\simeq 10^2~{\rm GeV}
\ee
should be satisfied, 
where $m_s \equiv s M_*$, 
$M \equiv m M_*$, $\bar{M} \equiv \bar{m} M_*$
$X(0) = x {\rm diag} (2,2,2,-3,-3)M_*$, 
$\bx(0) = \bar{x} {\rm diag} (2,2,2,-3,-3)M_*$, 
where $s$, $x$, $\bar{x}$, $m$ and $\bar{m}$ 
are dimensionless constants. 
Unlike Ref.~\cite{maru1}, 
we does not impose here that 
the mass of Higgs triplets should be above the GUT scale 
because this is not necessarily required 
to ensure the proton stability in our framework. 
%as will be discussed below. 
If we consider the case that 
the Higgs doublet and anti-doublets 
(triplet and anti-triplet) are localized 
at the same point for simplicity, 
then the condition (\ref{doublet}) is written as 
\bea
{\rm exp}[-\frac{1}{2}s(-6x + 2m - \frac{1}{2}s)] 
\simeq 10^{-16}. 
%\Leftrightarrow s(-6x + 2m - \frac{1}{2}s) 
%&\simeq& 16{\rm ln}10~
%{\rm for}~M_* \simeq 10^{18}~{\rm GeV}. 
\label{weak}
\eea
On the other hand, 
the mass of Higgs triplets $H_3$, $\bar{H}_3$ is 
\bea
M_3 \simeq M_* {\rm exp} 
\left[ - (2x + m)^2 + \frac{1}{4} (-s + 4x + 2m)^2 \right] 
\simeq 10^2 {\rm exp}(-5xs),
\eea
where Eq. (\ref{weak}) was used to obtain the last expression. 
Imposing $M_3 \ge 1{\rm TeV}$ leads to 
\be
\label{triplet}
-5xs \ge {\rm ln}10.
\ee 
One can easily check that there is a parameter region allowed 
by Eqs.~(\ref{weak}) and (\ref{triplet}) 
if $m^2 \ge \frac{77}{5}{\rm ln}10 \simeq 35.46$.

Since the Higgs doublets are localized at $y = (-3x+m)M_*^{-1}$ and 
the Higgs triplets are localized at $y = (2x+m)M_*^{-1}$, 
the relative distance between them is 5$|x| M_*^{-1}$. 
By adjusting $x$ appropriately, 
the baryon number violating dimension 5 operators 
are suppressed enough 
even if the Higgs triplets are light 
because Yukawa couplings of the Higgs triplet and 
the matter field localized around the Higgs doublets are small enough.
\footnote{The matter fields must be localized around the Higgs doublet 
to reproduce fermion masses and mixings.} 
For example, we consider the dimension five operators $QQQL$. 
Suppose that the relative distance between $Q$ 
and the Higgs doublets is $q$ in units of $M_*^{-1}$, 
and the relative distance between $L$ and the Higgs doublets 
is $l$ in units of $M_*^{-1}$. 
Assuming the Gaussian zero mode wave functions for the matter fields, 
one can write down the suppression factor of $QQQL$ as follows.
\footnote{Here we consider the case that $Q$ and $L$ are localized 
at the same side close to the Higgs triplets. 
The coefficients of $QQQL$ is much more suppressed 
in the case that either or both of $Q$ and $L$ are localized 
at the opposite side with respect to the Higgs doublets.} 
\bea
\frac{1}{M_3}~{\rm exp}~[-\frac{1}{3}~\{3(5|x|-q)^2 + (q-l)^2 
+ (5|x|-l)^2 \}] < 10^{-16}. 
\eea
The inequality is required to be consistent with 
the experimental data. 
Consider the case $M_3 \simeq {\rm TeV}$ and 
using the the typical solution in Ref.~\cite{maru1}, 
%taking into account 
%that $q$ and $l$ are at most $(5 \sim 6)$ 
%for the first gengeration, 
we obtain 
\bea
\label{pdecay}
|x| > 1.684. 
\eea
%
%Therefore, the proton decay via the dimension five operators 
%are suppressed if we adjust the parameter $x$ 
%satisfying the above condition (\ref{pdecay}). 
%We have checked that 
As for the dimension five operators $UUDE$, 
%are also suppressed enough 
%in the above range of parameters (\ref{pdecay}). 
the similar estimation tells us that 
$|x| > 1.653$ is enough for avoiding the rapid proton decay. 
Therefore, the proton decay via the dimension five operators 
are suppressed if we adjust the parameter $x$ 
satisfying the above conditions. 
We comment on the suppression of 
the dimension six baryon number violating operators. 
First, the dimension six operators by the X, Y gauge boson exchange 
are trivially suppressed since the masses of the X, Y gauge boson 
are the order of $10^{16}$ GeV. 
Second, the constraint for the dimension six operators 
by the Higgs scalar triplet exchange can be written as
\bea
\frac{1}{M_3^2}~{\rm exp}~[-\frac{1}{3}~\{3(5|x|-q)^2 + (q-l)^2 
+ (5|x|-l)^2 \}] < (10^{-16})^2. 
\eea
One can easily see that the bound (\ref{pdecay}) is enough to 
satisfy the above constraint 
because the upper bound of the exponential factor is 
${\rm exp}~[-25|x|^2] \simeq 10^{-30.8}$. 
In the light of this fact, 
the Higgs triplets with mass of order TeV is very interesting 
because we might be able to detect the Higgs triplets 
in collider experiments as an alternative signature of GUT 
even if the proton decay cannot be observed.

The second scenario is 
that the doublet-triplet splitting through the bulk Higgs mass term 
is acheived as a result of supersymmetry breaking, 
namely Giudice-Masiero mechanism \cite{GM}. 
Naively, 
GUT and Giudice-Masiero mechanism are incompatible 
because the light Higgs triplets necessarily appear 
and lead to the rapid proton decay. 
In our case, however, this is not true. 
Even if the Higgs triplets are light, 
the proton deccay is suppressed enough 
by natually small Yukawa couplings of the Higgs triplets 
to the matter. 
Let us assume that a singlet superfield $Z$ with nonvanishing F-term 
is localized on the brane at $y=0$.\footnote{In our scenario, 
the spectrum of a kind of the gaugino mediation \cite{gmed} is expected. 
Gauginos receive volume suppressed masses at the tree level 
since the wave functions of gaugino zero modes 
are flat in an extra dimension. 
Sfermions receive exponentially suppressed masses at the tree level 
since the wave functions of the matter zero modes are Gaussian 
and are localized on our wall apart from the brane at $y=0$.} 
We consider the following K\"ahler potential
\bea
&& \frac{1}{M_*} \int dy d^2\theta d^2\bar{\theta} 
~\delta(y)~(Z^\dag(x) H(x,y) \bar{H}(x,y) + {\rm h.c.}) \nonumber \\
&=& \frac{F_Z}{M_*} \phi_0(y=0) \bar{\phi}_0(y=0) 
\int d^2 \theta H_0(x)\bar{H}_0(x) + {\rm h.c.}.  
\eea
Substituting $\phi_0(y=0)$ and $\bar{\phi}_0(y=0)$ 
in (\ref{h}) and (\ref{hb}), 
the masses of the Higgs triplets and doublets are obtained as follows, 
\bea
M_3 &\simeq& \frac{F_Z}{M_*}~{\rm exp}[-(2x + m)^2], \\
M_2 &\simeq& \frac{F_Z}{M_*}~{\rm exp}[-(-3x + m)^2]. 
\eea
Here we assumed that the Higgs triplet (doublet) and 
anti-triplet (anti-doublet) localize at the same point, 
for simplicity. 
Requiring $M_2 \simeq M_W$, 
${\rm exp} [-(-3x+m)^2] \simeq M_*M_W/F_Z$ is obtained. 
This means $(-3x+m)^2 \simeq 2{\rm ln}10$ for 
$\sqrt{F_Z} \simeq 10^{11}$ GeV. 
In this case, the masses of the Higgs triplets become 
\bea
M_3 \simeq 10^4~{\rm exp}~[-(5x \pm \sqrt{2{\rm ln}10})^2]~{\rm GeV}. 
\eea
In order to be $M_3 \simeq {\rm TeV}$, $x \simeq -3.66, -0.628$. 
As is clear from the earlier discussion, 
%We have checked that 
it turns out that 
the dimension five and six operators are suppressed 
enough for $x \simeq -3.66$.

Finally, let us discuss the gauge coupling unification. 
%If the Higgs triplets are light, 
In our scenario with light Higgs triplets, 
the gauge coupling unification is lost. 
We can improve this point 
by simply introducing extra bulk matter fields 
${\bf 5}'$, ${\bf \bar{5}}'$ 
and by giving the GUT scale mass 
to the triplet components 
(denoted by ${\bf 3}'$ and ${\bf \bar{3}}'$) 
and the same mass as the Higgs triplets 
to the doublet components 
(denoted by ${\bf 2}'$ and ${\bf \bar{2}}'$). 
Yukawa couplings between these extra fields 
and ordinary chiral matter fields can be suppressed 
by the overlap of wave functions. 
The gauge coupling unification is preserved 
since ${\bf 2}'$ and ${\bf \bar{2}}'$ form a complete SU(5) 
multiplets with Higgs triplets.

We discuss how the above statement is realized. 
The action of the extra matter fields is similar to 
Eq.~(\ref{Higgsaction}) 
and the mass splitting is achieved by the bulk mass term 
\bea
\int d^4x dy \int d^2 \theta M_* {\bf 5}'(x,y) {\bf \bar{5}}'(x,y). 
\eea
After expanding in mode functions, 
one obtains masses of the ${\bf 3}', {\bf \bar{3}}'$ 
and ${\bf 2}', {\bf \bar{2}}'$ 
\bea
M_{3'} &\simeq& M_* {\rm exp} \left[ - \{(x' - \bar{x}') 
+ \frac{1}{2} (m' - \bar{m}')\}^2 \right] \gsim M_{GUT} 
\simeq 10^{16} {\rm GeV}, \\
M_{2'} &\simeq& M_* {\rm exp} \left[ - \{-\frac{3}{2}(x' - \bar{x}') 
+ \frac{1}{2} (m' - \bar{m}')\}^2 \right] = M_3, 
\eea
where $X' = x'{\rm diag}(2,2,2,-3,-3)M_*$, 
$\bx' = \bar{x}'{\rm diag}(2,2,2,-3,-3)M_*$, 
$M' \equiv m' M_*$ and $\bar{M}' \equiv \bar{m}' M_*$. 
$x'$, $\bar{x}'$, $m'$ and $\bar{m}'$ 
are dimensionless constants. 
These conditions are written as follows, 
\bea
&& -\sqrt{2 {\rm ln}10} \le 
(x'-\bar{x}') + \frac{1}{2} (m'-\bar{m}') \le 
\sqrt{2{\rm ln}10}, \\
&& -3(x' - \bar{x}') + (m' - \bar{m}') 
\simeq \pm \sqrt{s(2x + m - s/4)}. 
\eea
One can easily see that 
these two conditions are satisfied 
%by ${\cal O}(1)$ parameters $x', \bar{x}', m'$ and $\bar{m}'$ 
($x'-\bar{x}' = 3, m' - \bar{m}' = -3$ for example.).

In summary, 
we have discussed the possibility 
that the Higgs triplet can be light 
without contradictions with the proton stability 
%and the gauge coupling unification 
in the context of higher dimensional theory. 
The proton stability is ensured 
by the suppression of Yukawa coupling 
of the Higgs triplet to the matter 
through its small overlap of wave functions 
in extra dimensions. 
%The gauge coupling unification is preserved 
%by introducing extra bulk matter fields and 
%causing the mass splitting 
%so that the doublet components 
%form a complete SU(5) multiplets with the Higgs triplets and 
%the triplet components become superheavy. 
%These two operations are carried out by at most ${\cal O}(10)$ 
%tuning of parameters in the fundamental theory 
%in contrast to an unnatural fine-tuning of parameters 
%in four dimensional case. 
Phenomenologically interesting is that 
Higgs triplets with mass of order TeV might be detected 
in future collider experiments 
as an alternative signature of GUT 
instead of the proton decay. 
The gauge coupling unification can be preserved 
by introducing extra bulk matter fields and 
causing the mass splitting 
so that the doublet components 
form a complete SU(5) multiplets with the Higgs triplets and 
the triplet components become superheavy.

%%%%%%%%%%%%%%%%%%%%%%%%%%%%%%%%%%%%%%%%%%%%%%%%%%%%%%%%%%%%%%%%%%%%%%%%%%%%%%%
%                             Acknowlegdements                                %
%%%%%%%%%%%%%%%%%%%%%%%%%%%%%%%%%%%%%%%%%%%%%%%%%%%%%%%%%%%%%%%%%%%%%%%%%%%%%%%

\begin{center}
{\bf Acknowledgments}
\end{center}
We would like to thank the organizers of the workshop ``post NOON" 
where this work was initiated. 
N.H. is supported by the Grant-in-Aid for Scientific Research, 
Ministry of Education, Science and Culture, Japan (No.12740146) and 
N.M. is supported 
by the Japan Society for the Promotion of Science 
for Young Scientists (No.08557).

%%%%%%%%%%%%%%%%%%%%%%%%%%%%%%%%%%%%%%%%%%%%%%%%%%%%%%%%%%%%%%%%%%%%%%%%%%%%%%%
%                                  References                                 %
%%%%%%%%%%%%%%%%%%%%%%%%%%%%%%%%%%%%%%%%%%%%%%%%%%%%%%%%%%%%%%%%%%%%%%%%%%%%%%%

\end{document}